\documentclass[tikz,aps,pra,twocolumn,superscriptaddress]{revtex4}
\usepackage[usenames,dvipsnames]{color}
\usepackage{tikz}
\usepackage{graphicx}
\usepackage{dcolumn}
\usepackage{bm}
\usepackage{amsmath}
\usepackage{amsfonts}
\usepackage{psfrag}
\usepackage[bookmarks,colorlinks=false]{hyperref}
\usepackage{mathtools}
\usepackage{accents}

\usepackage[mathcal]{euscript}

\usetikzlibrary{arrows,intersections,decorations.pathmorphing}

\def\bra#1{\mathinner{\langle{#1}|}} 
\def\ket#1{\mathinner{|{#1}\rangle}}

\newcommand{\Eq}[1]{Eq.~(\ref{#1})}
\newcommand{\Eqs}[2]{Eqs.~(\ref{#1})-(\ref{#2})}

\begin{document}

\title{Strong coupling of ionising transitions}

\author{Erika Cortese}
\affiliation{School of Physics and Astronomy, University of Southampton, Southampton, SO17 1BJ, United Kingdom}
\author{Iacopo Carusotto}
\affiliation{INO-CNR BEC Center and Dipartimento di Fisica, Universita di Trento, I-38123 Povo, Italy}
\author{Raffaele Colombelli}
\affiliation{Centre de Nanosciences et de Nanotechnologies, CNRS UMR 9001, Universit\'e Paris-Sud, Universit\'e Paris-Saclay, C2N - Orsay, 91405 Orsay cedex, France}
\author{Simone \surname{De Liberato}}
\affiliation{School of Physics and Astronomy, University of Southampton, Southampton, SO17 1BJ, United Kingdom}

\begin{abstract}
We demonstrate that a ionising transition can be strongly coupled to a photonic resonance. The strong coupling manifests itself with the appearance of a narrow optically active resonance below the ionisation threshold. Such a resonance is due to electrons transitioning into a novel bound state created by the collective coupling of the electron gas with the vacuum field of the photonic resonator. Applying our theory to the case of bound-to-continuum transitions in microcavity-embedded doped quantum wells, we show how those strong-coupling features can be exploited as a novel knob to tune both optical and electronic properties of semiconductor heterostructures.
\end{abstract}

\maketitle

\section{Introduction}
When a single photon can be trapped long enough in an optical resonator to undergo multiple absorption and re-emission cycles, the coupled light-matter system is said to be in the strong coupling regime. Its physics cannot then be correctly described in terms of irreversible absorption and emission of photons, but it becomes necessary to consider instead hybrid quasiparticles, half-light half-matter, named polaritons \cite{Kavokin}.

Many works demonstrated how the hybridization with matter strongly alters not only the spectrum, but also the field profile \cite{DeLiberato14,Bayer17,Passler18}, and the quantum \cite{Savasta05,LeBoite16,Garziano17,Munoz14} and nonlinear \cite{Carusotto13,Gubbin17} properties of the photonic resonator.
More recently interest has also broadened to investigate how strong coupling can be used to modify properties of the underlying matter degrees of freedom \cite{Galego15,Cwik16,Cortese17,Luk17,Flik17,Keeling18,Citrin03,Brodbeck2017},   including changes in electrical \cite{Orgiu15,Feist15,Hagenmuller18,Ciuti18,Paravicini18} and photochemical \cite{Hutchison12,Hutchison13,Herrera16,MartinezMartinez18} properties.

\begin{figure}[htbp]
\begin{center}
\includegraphics[width=9cm]{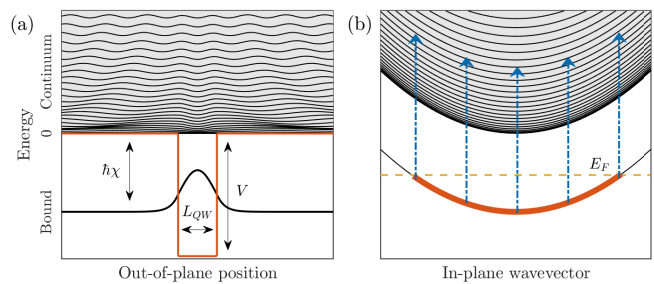}
\label{Fig1}
\caption{{Schematic representation of the electronic structure of a quantum well of width $L_{QW}$, with a single bound state below the continuum (shaded in gray). (a) Electronic density envelope functions of the different single-particle eigenmodes, shifted by their energy. The potential profile is plotted in red, with the first ionisation energy $\hbar\chi$ and the conduction band discontinuity $V$ explicitly marked.
The figure has been obtained using the parameters of the structure described in Sec. \ref{SingleQW}. For sake of clarity only one continuum mode every ten is shown, with its density multiplied by ten.  (b) In-plane dispersion of the different subbands in momentum space. The single bound state is filled with electrons (in red), up to the Fermi energy $E_F$ (dashed yellow line). The dash-dotted blue arrows represent a collective bound-to-continuum transition.}}
\end{center}
\end{figure}
Notwithstanding a large and quickly growing interest in polaritonic physics, until now only transitions between bound states have been exploited for achieving strong coupling. Bound-to-continuum, ionising transitions, would in fact seem to be by their very nature irreversible, as the ionised components get separated upon photon absorption, thus not allowing for a subsequent re-emission.

In this work we demonstrate that ionising electronic transitions can be strongly coupled to a photonic resonator. Such a  bound-to-continuum strong coupling manifests itself with the appearance of a narrow polaritonic resonance below the ionising threshold. This suggests the concomitant appearance of a novel, vacuum-induced bound electronic state, not present in the uncoupled Hamiltonian. 
Apart from its importance from a fundamental perspective, this discovery opens the way to a number of practical applications as those novel bound states can be pathways for chemical reactions, improve the efficiency of multi-photon transitions, or push the tunability of semiconductor devices beyond what allowed by the mere electronic properties.

Although most of the concepts introduced in this paper are general enough to be broadly applied to any cavity quantum electrodynamics platform, including atomic and molecular systems, for the sake of definiteness henceforth we will  consider the specific case of microcavity-embedded doped quantum wells, sketched in Fig. \ref{Fig1}. 
In those systems the confinement along the growth ($z$) axis splits the conduction band into multiple discrete bound or continuous unbound subbands. The parabolic quasi-parallel in-plane dispersion then allows for the excitation of resonant coherent electronic transitions with long dephasing times. 
In the case in which more than one bound subband is present, the resulting narrow optical transition has been successfully strongly coupled with mid-infrared and THz resonators \cite{Dini03,Anappara09,Todorov10}. The resulting quasiparticles, named intersubband polaritons, have been highlighted as a promising platform for long-wavelength optoelectronics \cite{DeLiberato09,DeLiberato13,Colombelli15,Manceau18}, with the possibility to tune the doping {\it in-situ} by optical or electrical means \cite{Anappara05,Guenter09}.
The role of intersubband polariton formation on the electronic wavefunction has been investigated in the past \cite{DeLiberato09b}, highlighting the possibility to dramatically increase the emission efficiency by injecting electrons into superradiant bright states. Bound-to-continuum transitions in doped quantum wells have also been the object of theoretical \cite{Zaluzny91,Zaluzny97} and experimental \cite{Capasso92} investigations.

\section{Theory}
In this Section we will sketch the theory of the bound-to-continuum strong coupling and the calculation of the corresponding vacuum-induced bound electronic state wavefunctions. Detailed derivations can be found in the the Appendix.

\subsection{Spectrum}
The problem of the coupling between a discrete resonance (in our case the photonic mode) and a continuum (the bound-to-continuum electronic transitions) was initially treated by Fano in his landmark paper \cite{Fano56}. In such a work the coupled eigenfrequencies are always assumed to fall into the uncoupled continuum, leading to the characteristic asymmetric broadened absorption lineshape.
The limit of a very narrow continuum, describing an inhomogeneously broadened discrete resonance, has also been investigated \cite{Houdre96,Manceau17}.
Here we will consider instead the case of a semi-infinite ionisation continuum of unbound states, in which one of the hybridised light-matter eigenmodes lies below its lowest edge. This can happen either because the uncoupled discrete resonance is not resonant with the continuum to start with, or because the coupling is large enough to {\it push} a coupled eigenmode out of the continuum. 

In a planar semiconductor heterostructure the electronic states can be indexed by the in-plane two-dimensional momentum vector $\mathbf{k}$ and by an index $n$ which runs over both bound and continuum states quantised along the growth, out-of-plane axis. The field operator for electrons of in-plane momentum $\mathbf{k}$ can thus be written as 
\begin{eqnarray}
\Psi_{\mathbf{k}}(z)&=&\sum_n \phi_{n}(z)c_{n\mathbf{k}},
\end{eqnarray}
where  $c_{n\mathbf{k}}$ is the fermionic annihilation operator of the $n^{\text{th}}$ electron level with in-plane wavevector $\mathbf{k}$, frequency $\omega^c_{n\mathbf{k}}$, and envelope wavefunction $\phi_{n}(z)$. The electronic single-particle spectrum will be characterised by one or more bound levels $\omega^c_{n\mathbf{k}}<0$, and a series of unbound states delocalised across the bulk $\omega^c_{n\mathbf{k}}>0$. Those states can either form a real continuum or arrange themselves in a set of broadened minibands, depending on the details of the potential felt 
by the electrons. We will in the following use the term {\it continuum} with the understanding that it can cover both situations. 

Collective electronic transitions between single-particle states  can be indexed by the in-plane wavevector $\mathbf{q}$ and the index $\alpha\equiv(nm)$, with $m$ and $n$ respectively the initial and final electronic levels. Specialising to the case of interest here we consider a system in which only bound-to-continuum transitions are optically active, by choosing the number of electrons $N$ such that, as depicted in Fig. \ref{Fig1} (b), the Fermi energy stands between the bottom of the last bound and first unbound subbands.

A cavity photon with in-plane wavevector $\mathbf{q}$ thus couples to electronic transitions described by the dipole operators
\begin{eqnarray}
b^{\dagger}_{\alpha\mathbf{q}}&=&\frac{1}{\sqrt{N}}\sum_{\mathbf{k}}c_{n\mathbf{k+q}}^{\dagger}c_{m\mathbf{k}}.
\end{eqnarray}
In the dilute excitation regime, in which the number of excitations in the system is much smaller than $N$, those operators satisfy bosonic commutation relations \cite{Ciuti05,DeLiberato09,Shammah17}
\begin{eqnarray}
\left[ b_{\alpha\mathbf{q}},\, b^{\dagger}_{\beta\mathbf{q'}} \right]&=&\delta_{\alpha\beta}\delta(\mathbf{q-q'}).
\end{eqnarray}
Many-body plasmonic effects can become important at high doping densities. A Bogoliubov transformation then allows to express the single particle transitions $b_{\alpha\mathbf{q}}$  as superpositions of different multisubband plasmon modes $p_{\alpha \mathbf{q}}$ \cite{DeLiberato12,Todorov12,Todorov15}
\begin{eqnarray}
(b^{\dagger}_{\alpha \mathbf{q}}+b_{\alpha\mathbf{-q}})&=&\sum_{\beta}h_{\alpha \beta}(p_{\beta \mathbf{q}}^{\dagger}+p_{\beta \mathbf{-q}}).
\end{eqnarray}
Introducing $a_{\mathbf{q}}^{\dagger}$, the bosonic creation operator for a cavity photon of in-plane wavevector $\mathbf{q}$ and frequency $\omega^a_{\mathbf{q}}$, the light-matter Hamiltonian takes the form
\begin{eqnarray}
\label{HAP}
H&=&\sum_{\mathbf{q}}\left[\hbar\omega^a_q a^{\dagger}_{\mathbf{q}}a_{\mathbf{q}}+\sum_{\alpha} \hbar\omega^p_{\alpha}p_{\alpha\mathbf{q}}^{\dagger}p_{\alpha\mathbf{q}}
\right.\\ &&\left.\nonumber+ 
\sum_\alpha \frac{\hbar \Xi_{\alpha q}}{2}
 (a^{\dagger}_{\mathbf{-q}}+a_{\mathbf{q}})
(p_{\alpha\mathbf{q}}^{\dagger}+p_{\alpha\mathbf{-q}})\right],
\end{eqnarray}
where $\omega^p_{\alpha}$ are the frequencies of the electronic transitions dressed by local-field effects and $\Xi_{\alpha q}$ is the renormalised light-matter coupling.
This Hamiltonian can be diagonalised in term of bosonic hybrid light-matter polariton operators
\begin{eqnarray}
\label{polop}
d_{s\mathbf{q}}=x_{sq} a_{\mathbf{q}}+z_{sq} a^{\dagger}_{\mathbf{-q}}+
\sum_{\alpha} \left[ y_{s \alpha q}p_{\alpha \mathbf{q}}+w_{s \alpha q}p_{\alpha \mathbf{-q}}^{\dagger}\right].
\end{eqnarray}
When the index $s$ runs over solutions which are in the continuum, the system is equivalent to the one described by Fano and we will not explicitly discuss its solution here.
We consider instead solutions with frequency $\omega^d_{sq}<\chi$, where we defined the frequency of first ionisation $\chi$ as the lowest $\omega^p_{\alpha}$ belonging to a continuum part of the spectrum (see Fig. \ref{Fig1} (a)). In the Appendix it is shown that in this case the eigenvalue $\omega^d_{sq}$ will obey the  eigenequation
\begin{eqnarray}
\label{eigenw}
\frac{\omega^a_{q}}{\omega_{q}^{a2}-\omega^{d2}_{sq}}\sum_{\alpha}
\frac{\lvert \Xi_{\alpha q}\rvert^2\omega^p_{\alpha}}{\omega_{\alpha}^{p2}-\omega^{d2}_{sq}}&=&1.
\end{eqnarray}
The existence of a solution satisfying \Eq{eigenw} can be easily proved if the photonic mode lies below the frequency of first ionisation ($\omega^a_q<\chi$), but in the opposite case ($\omega^a_q>\chi$) one needs to solve the integral equation to verify whether the interaction is strong enough to {\it push} the hybridised polariton mode out of the continuum.

\begin{figure}[htbp]
\begin{center}
\includegraphics[width=9.5cm]{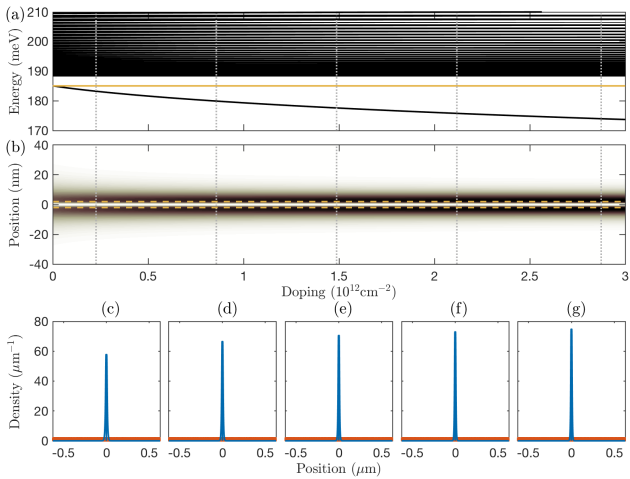}
\caption{\label{Fig2185} Simulation of a single quantum well of width $L_{QW}=4$nm in a bulk of total width $T=1\mu$m, with effective cavity length $L_c=25nm$ and cavity energy $\hbar\omega^a_q=185$meV, 
smaller than the first ionisation $\hbar\chi=188.4$meV. 
(a) Polaritonic spectrum as a function of the doping. The yellow line marks the cavity energy. (b) Colormap of the excited electron density $\lvert \psi^e_{sq}(z)\rvert^2$ for the lowest lying polaritonic mode as a function of doping. Yellow dashed lines mark the boundaries of the quantum well. (c-g) Plots of $\lvert \psi^e_{sq}(z)\rvert^2$ for all the polaritonic modes (all the  $s$ up to the cutoff) relative to the five values of doping marked by dotted vertical grey lines in panels (a,b).
The lowest lying mode represented in panel (b) is plotted in blue, all the other modes in the continuum are instead coloured in red, forming the thin homogeneous red band of density $T^{-1}$ visible at the bottom of each panel. Note that, due to the different scale, the node of the lowest lying localised electronic density visible in panel (b), is not clearly resolved in panels (c-g). 
}
\end{center}
\end{figure}

\begin{figure}[htbp]
\begin{center}
\includegraphics[width=9.5cm]{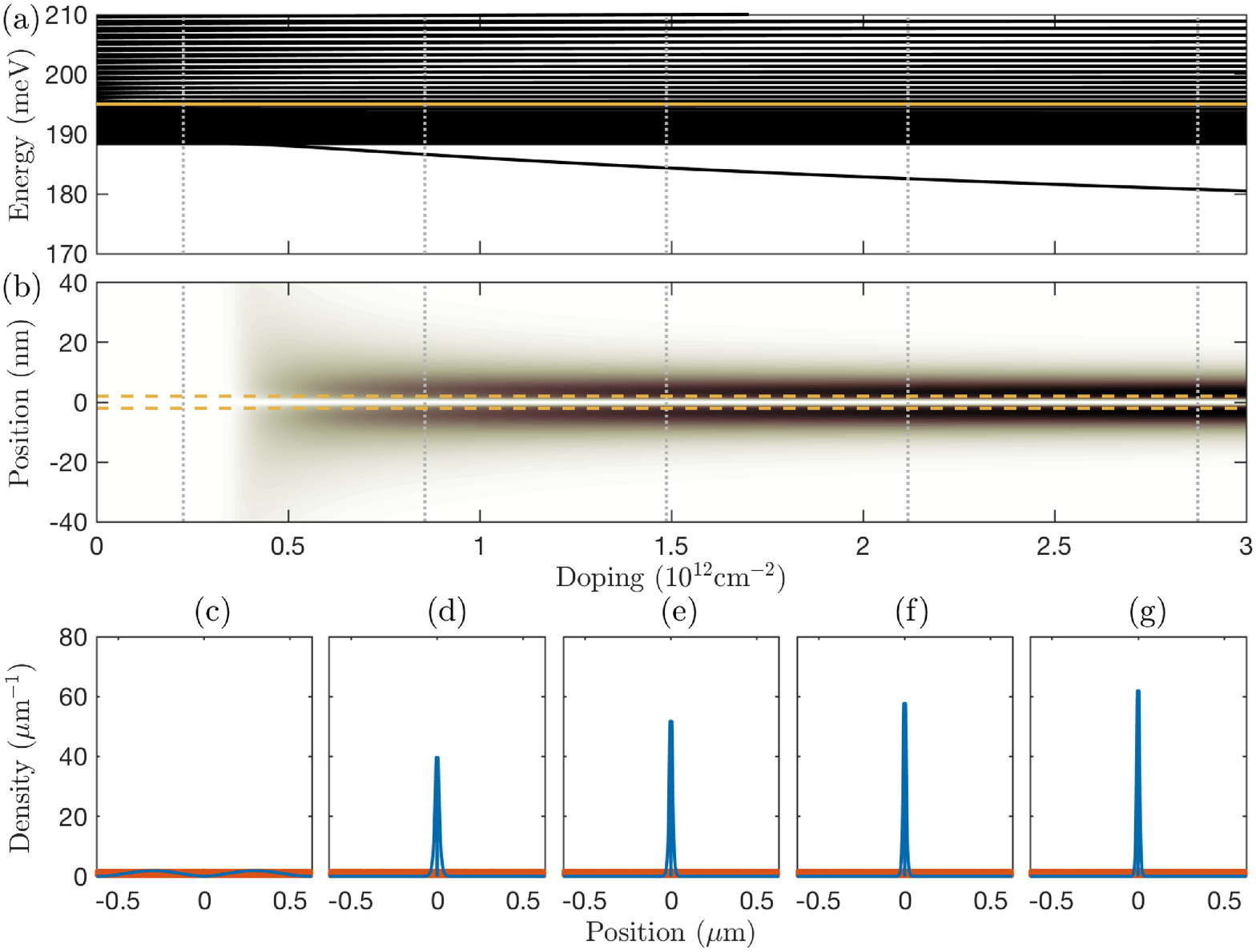}
\caption{\label{Fig2195} Same as in Fig. \ref{Fig2185} but with cavity energy $\hbar\omega^a_q=195$meV, larger than 
the first ionisation $\hbar\chi=188.4$meV.} 
\end{center}
\end{figure}

\begin{figure}[htbp]
\begin{center}
\includegraphics[width=9.5cm]{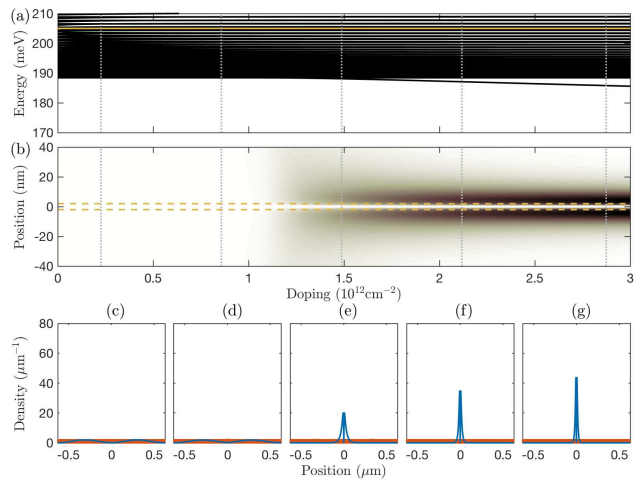}
\caption{\label{Fig2205} Same as in Fig. \ref{Fig2185} but with cavity energy $\hbar\omega^a_q=205$meV, larger than 
the first ionisation $\hbar\chi=188.4$meV.}
\end{center}
\end{figure}

\subsection{Electronic density}
Assuming for the moment that \Eq{eigenw} has at least a solution $\omega^d_{sq}<\chi$, we will here study its properties. The operator in \Eq{polop} describes the normal mode of the system as a superposition of photonic and matter excitations but, notwithstanding the fact all electronic transitions are in the continuum, and thus all the available final states unbound, there is not enough energy to promote an electron to an unbound state. The particular linear superposition of delocalised states specified by the $y_{s \alpha q}$ and $w_{s \alpha q}$ coefficients in \Eq{polop} has thus to describe transitions to a bound state non present in the uncoupled Hamiltonian. In order to visualise such a state we can define the electron density
\begin{eqnarray}
N(z)&=&\sum_{\mathbf{k}}\Psi_{\mathbf{k}}^{\dagger}(z)\Psi_{\mathbf{k}}(z),
\end{eqnarray}
and calculate its difference between the ground state $\ket{G}$ and the state with one excitation in an arbitrary $d_{s\mathbf{q}}$ polaritonic mode
\begin{eqnarray}
\label{DN}
\Delta N_{s\mathbf{q}}(z)&=&\bra{G}d_{s\mathbf{q}} N(z)d^{\dagger}_{s\mathbf{q}}\ket{G}-\bra{G}N(z) \ket{G}\\&=&P_{s{q}}\left[\lvert \psi^e_{s{q}}(z) \rvert^2-\lvert \psi^g_{s{q}}(z) \rvert^2\right],\nonumber
\end{eqnarray}
where, as detailed in the Appendix,
$P_{s{q}}$ is the weight of the matter component of the polaritonic mode, and  $\psi^g_{s{q}}(z)$ and $\psi^e_{s{q}}(z)$, built respectively only by wavefunctions of initially full and empty electronic levels, naturally lend themselves to be interpreted as the initial (ground) and final (excited) wavefunctions of the electronic transition generated by the operator $d_{s\mathbf{q}}$. 

\section{Results}
We will now apply the previously developed theory to the case 
of $n_{QW}$ identical doped GaAs/Al$_x$Ga$_{1-x}$As quantum wells
of width $L_{QW}=4$nm. The Al fraction $x=0.33$ then assures to have a single bound state per quantum well. In those structures intersubband polaritons have been demonstrated with doping levels up to $N_{2DEG}=3\times10^{12}$cm$^{-2}$ \cite{Manceau18}. Electronic bare wavefunctions are then calculated considering a single conduction band in the effective mass approximation and using Bastard boundary conditions for the envelope functions \cite{Bastard}. These eigenfunctions are used to calculate all the electronic parameters of \Eq{HAP}. The resulting multi-mode Hopfield matrix is then numerically diagonalised, leading to the determination of the eigenfrequencies  $\omega^d_{sq}$ and of the eigenvectors in \Eq{polop}. Those coefficients are then used in \Eq{DN} to calculate the ground and excited electronic densities $\lvert\psi^g_{sq}(z)\rvert^2$ and $\lvert\psi^e_{sq}(z)\rvert^2$, and the matter weight $P_{sq}$.

\subsection{$n_{QW}=1$}
\label{SingleQW}
In order to build a good understanding of the implications of bound-to-continuum strong coupling we start by considering a single quantum well in a wide bulk of total width $T=1\mu$m, modeling well a ionisation continuum in which the electron can escape and be lost. The envelope functions in Fig. \ref{Fig1} (a) have been calculated using those parameters. 
In order to get large values of the coupling with a single quantum well we consider the electronic transition coupled to a subwavelength photonic resonance with an effective cavity length $L_c$ much smaller than the transition wavelength $\lambda$. In particular we fix
$L_c=25$nm, which for a photonic transition quasi-resonant to the frequency of first ionisation $\chi$, implies $\frac{\lambda}{2L_c}\simeq 125$, a value which can be today achieved using various mid-infrared architectures \cite{Benz13,Caldwell13,Gubbin17b}. Note that specific implementations could impose further constraints on the shape of the bulk ({\it e.g.}, forcing the quantum well to be at a specific distance from one of the boundaries), but for the sake of definiteness here we will neglect this possibility and consider the quantum well to be placed in the center of the $1\mu$m bulk. A microscopic model for the photonic resonator is instead considered for the case of multiple quantum wells described below.

In Figs. \ref{Fig2185},\ref{Fig2195}, and \ref{Fig2205} we show the results for cavity energies $\hbar\omega^a_q=185,195$, and $205$meV respectively. In panel (a) of each figure we plot the polaritonic spectrum as a function of the doping, clearly showing the ionisation continuum around $\hbar\chi=188.4$meV. A single discrete mode below the continuum is also present, for any doping in the case $\omega^a_q<\chi$, and only above a critical doping for $\omega^a_q>\chi$, as in this case the coupling needs to be strong enough to push a polaritonic mode out of the continuum. In panels (b) we plot instead the normalised excited state density $\lvert \psi^e_{sq}(z) \rvert^2$ for the lowest-lying polaritonic state in a neighbourhood of the quantum well.
Consistently with our interpretation a localised electronic mode forms only when a discrete polaritonic resonance is present.
Note that such a resonance, not having enough energy to decay in the ionisation continuum, is expected to have a linewidth of the order of few meV, determined by non-parabolicity, electron-phonon, and electron-electron scattering \cite{Waldmuller04}, as well as by the coupling of its photonic component to extra-cavity radiative modes. The predicted coupling-induced shifts are thus sizeably larger than the expected linewidths, making those resonances spectroscopically observable and individually addressable.

\begin{figure}[htbp]
\begin{center}
\includegraphics[width=8cm]{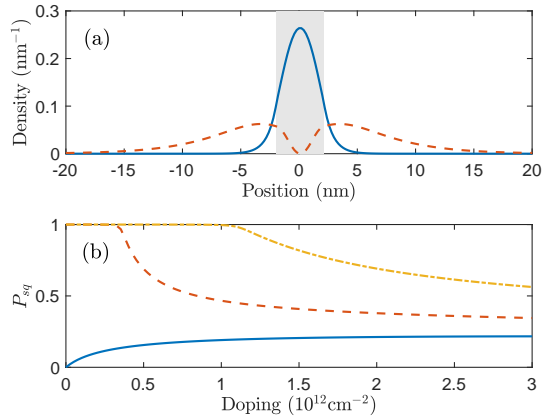}
\caption{\label{FigPM} (a)  Ground (blue solid line) and excited (red dashed line) electronic densities corresponding to the lowest eigenmode in Fig. \ref{Fig2195} (g). The shaded region corresponds to the quantum well.
(b) Weight of the matter component $P_{sq}$ for the lowest-lying polaritonic mode as a function of doping for $\hbar\omega^a_q=185$meV (blue solid line), $\hbar\omega^a_q=195$meV (red dashed line), and $\hbar\omega^a_q=205$meV (yellow dash-dotted line). Other parameters as in Fig. \ref{Fig2185}.}
\end{center}
\end{figure}

In order to prove that excited localised electronic states do not exist in the continuum below the critical density, in panels (c)-(g) we plot the excited electronic density $\lvert \psi^e_{sq}(z)\rvert^2$ for all the polaritonic modes, using a cutoff of  $500$meV on the single electron energies.
Lines corresponding to all values of $s$ except the lowest-lying one are all plotted in red, and their overlap forms the uniform red band which can be seen on the very bottom of each panel, of homogeneous density $T^{-1}$.  Such homogeneous density, vanishing in the limit $T\rightarrow\infty$ of a true continuum, is what expected from continuum of delocalised modes. 
The density of the first mode is instead plotted in blue. Those results further confirm that a localised electronic mode emerges only when a discrete polaritonic mode appears. Note that due to the different scales of panel (b) and (c)-(g) the node of the excited density is not clearly visible in the latter. 
In panel (a) of Fig. \ref{FigPM} we plot both the ground state density $\lvert \psi^g_{sq}(z)\rvert^2$  (blue solid line), which doesn't depend either on $s$ or on doping, and the excited density $\lvert \psi^e_{sq}(z)\rvert^2$ (red dashed line) corresponding to the lowest eigenmode in Fig. \ref{Fig2195} (g) in a $40$nm interval around the quantum well.
From such a figure we can both verify that the ground state density $\lvert \psi^g_{sq}(z)\rvert^2$ defined in \Eq{DN} corresponds to the initially occupied electronic state, and that the excited state density is compatible with a localised odd-symmetry wavefunction.

Of course the normalised electronic densities $\lvert \psi^e_{sq} \rvert^2$ only give a partial information on the existence and observability of vacuum-induced bound electronic states, because from \Eq{DN} they are weighted by the matter fraction $P_{sq}$. In Fig. \ref{FigPM} (b) we thus plot 
$P_{sq}$ relative to the mode plotted in panels (b) of Figs. \ref{Fig2185},\ref{Fig2195}, and \ref{Fig2205} as a function of doping, for the three considered values of the cavity energy. As expected, when the bare cavity mode is below the continuum ($\hbar\omega^a_q=185$meV, blue solid line), for vanishing doping the discrete lowest-lying polaritonic mode is just the bare cavity with a vanishing matter component. In the opposite cases ($\hbar\omega^a_q=195$meV, red dashed line, and $\hbar\omega^a_q=205$meV yellow dash-dotted line) the lowest mode is initially purely matter ($P_{sq}=1$) and only when the discrete polaritonic mode appears we observe light-matter hybridisation. 
In the three cases though, a strong hybridisation is observed for
experimentally achievable values of the doping.

\begin{figure}[htbp]
\begin{center}
\includegraphics[width=9.5cm]{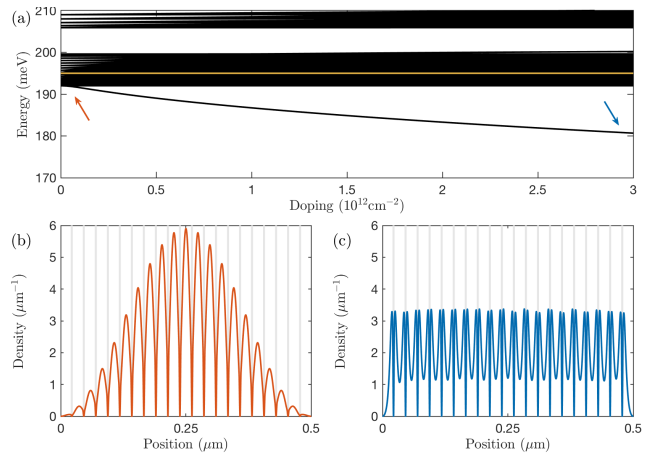}
\caption{\label{Fig3} Simulation of $n_{QW}=20$ quantum wells of width $L_{QW}=4$nm in a bulk of total width $T=0.5\mu$m embedded in a planar microcavity. The  TM0 mode of the microcavity is chosen to have energy $\hbar\omega^a_q=195$meV, larger than the first ionisation energy $\hbar\chi=192$meV. 
(a) Polaritonic spectrum as a function of the doping. The yellow line marks the cavity energy. (b,c) Plots of the excited electron density $\lvert \psi^e_{sq}(z)\rvert^2$ for the lowest lying polaritonic mode, for values of doping equal respectively to $N_{2DEG}=0$ and $3\times10^{16}$cm$^{-2}$, marked by arrows in panel (a). Shaded regions correspond to the locations of the quantum wells.} 
\end{center}
\end{figure}

\begin{figure}[htbp]
\begin{center}
\includegraphics[width=9cm]{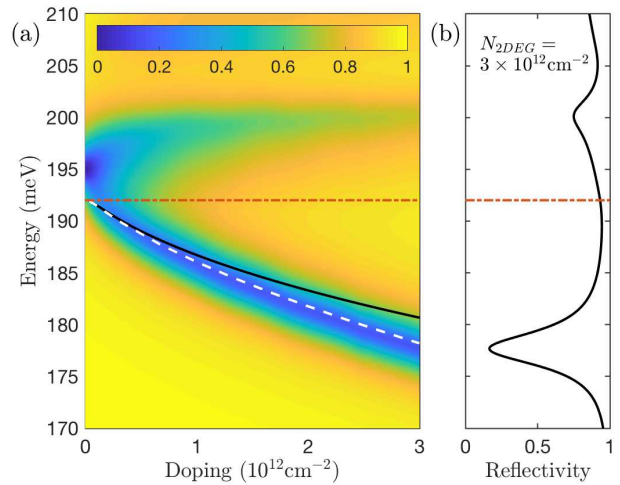}
\caption{\label{Fig4} (a) Reflectivity map for the same structure studied in Fig. \ref{Fig3}, calculated considering an electronic linewidth of $4$meV.
The horizontal dash-dotted red line marks the first ionisation energy. The solid black and dashed white lines mark instead the dispersion of the lowest polariton mode obtained using the Hopfield approach respectively without and with the effective medium approximation. (b) A vertical cut of panel (a) for $N_{2DEG}=3\times 10^{12}$cm$^{-2}$.}
\end{center}
\end{figure}

\subsection{$n_{QW}=20$}
After having investigated the single quantum well case, and demonstrated the formation of discrete electronic states out of the continuum for suitable strength of the light-matter interaction, here we will consider the case of $n_{QW}=20$ quantum wells which couple to the TM0 mode of a planar resonator as wide as the electron bulk. The relative facility to achieve strong coupling in multiple-quantum well structures \cite{Dini03,Anappara09,Todorov10} will allow for a rapid observation of the discrete polaritonic resonances emerging out of the continuum, heralding the generation of vacuum-induced bound electronic states.

We consider a sample with $n_{QW}=20$ quantum wells, separated by barriers of $20$nm, with the same length $L_{QW}=4$nm used previously, and the bulk and the resonator have a total length $L_c=T=0.5\mu$m. 
The filling factor $n_{QW}L_{QW}/L_c=0.16$, which in Ref. \cite{Todorov12} was shown to be the relevant figure of merit to quantify the strength of the light-matter interaction is thus the same as in Sec. \ref{SingleQW}, allowing for a meaningful comparison.

In Fig. \ref{Fig3} (b) we plot the eigenmodes for $\hbar\omega^a_q=195$meV. Notwithstanding minor differences, including 
a larger energy of first ionisation $\hbar\chi=192$meV,
which are to be expected given the different structure of the continuum, we recover results in agreement with those in 
Fig. \ref{Fig2195} (a), showing that discrete resonances below the continuum are observable also in multi-quantum well structures. 
In panels (b) and (c) we plot the excited electronic density in the lowest-lying eigenmode, for values of the doping $N_{2DEG}=0$ and $3\times 10^{12}$cm$^{-2}$ respectively, marked by arrows in panel (a).
It can be noticed that the electronic densities for the continuous (red) and discrete (blue) modes are not as spectacularly different as in Fig. \ref{Fig2195}.
This was expected because in a multi-quantum wells structure ionised electrons will still be confined in the proximity of quantum wells, as the total space occupied by the quantum wells is now a macroscopic fraction of the total width of the bulk. Nevertheless, we can recognise the same physics at play, as the collective density, although modulated by the presence of the quantum wells, passes from a mode of the entire structure, whose intensity is maximal at the center, to a localised one in which the wavefunction is roughly the same in each quantum well.

Finally, with the objective to test the accuracy of our quantum theory and to provide an experimentally accessible observable, we calculated the reflectivity of the same structure coupled to a planar metallic resonator, using a classical transfer matrix approach in the effective medium approximation including local field effects \cite{Zaluzny91,Zaluzny97,Zaluzny99} and considering an electronic linewidth of $4$meV. Results are shown in Fig. \ref{Fig4}, which replicates acceptably well the modal spectrum of Fig. \ref{Fig3} (a).
Superimposed on the reflectivity map we plot with a solid black line the lowest lying polariton from \ref{Fig3} (a). The small difference between the Hopfield and transfer matrix approaches, visible in the increased red-shift of the reflectivity dip corresponding to the discrete polariton mode in the transfer matrix results, is due to the use of an effective medium approximation. We verified the origin of such discrepancy by implementing the same approximation in the Hopfield model, as described in Ref.  
\cite{Todorov12} and explained in the Appendix, leading to the dashed white line which instead correctly follows the reflectance dip up to high levels of doping.

\section{Conclusions}
In this work we demonstrated that a ionising electronic transition can be strongly coupled to a photonic resonator. Spectroscopically this strong coupling manifests itself with the appearance of a discrete optical resonance below the ionisation threshold. This strongly suggests that such a resonance must be accompanied by the appearance of a novel bound electronic state.

The natural next question then is how the existence of these states can be 
directly probed. From the optical side, the width and the shape of the 
resonance bears the typical signatures of localization. For instance the 
absorption of a bound-to-continuum transition has a large and asymmetric 
shape, while a bound-to-bound transition is narrower (typically below 
10\% at room temperature) and Lorentzian-shaped (see for example 
Ref. \cite{Capasso92}). However, a truly direct way to prove the electronic localization effect would require electrical measurements. One possibility is to use scanning tunnel microscopy. An interesting alternative, that can open up vaster perspectives in the long term, is to develop quantum well (QWIP) 
or quantum cascade (QCD) intersubband detectors operating in strong 
coupling. Such devices rely on polaritonic excitations tunneling out 
into electronic states: the study of the transport in such devices can 
directly prove an electronic localization effect.  

Our results can find direct applications in the physics and technology of intersubband transitions in doped quantum wells, where the potential of exploiting bound-to-continuum transitions \cite{Faist01,Scalari05} and of cavity-induced modifications of electronic states  \cite{DeLiberato09b}  were already recognised.
More broadly this investigation, which could be extended to other cavity quantum electrodynamics systems both in solid-state and atomic physics, demonstrates a novel way strong coupling can be exploited to  influence the properties of materials coupled to light, with potential impact on fields ranging from chemistry to material science.

\section{Acknowledgments}
S.D.L. is a Royal Society Research Fellow and acknowledges support from the Innovation Fund of the EPSRC Programme EP/M009122/1. 
R.C. acknowledges support from the European Research Council (IDEASERC) (``GEM'') (306661).
R.C. and I.C. acknowledge support from the European Union FET-Open Grant MIR-BOSE 737017.

\appendix

{\widetext
\section{Calculation of the bare electronic wavefunctions}
Following the approach described in Ref. \cite{Bastard} we determine the electronic envelope functions $\phi_n(z)$ and their subband-edge energies $\hbar \omega^c_{n\mathbf{0}}$ 
numerically solving the eigenequation 
\begin{eqnarray}
\label{BDD}
\left[-\frac{\hbar^2}{2}\frac{\partial}{\partial z}\frac{1}{m^*(z)}\frac{\partial}{\partial z}+V(z)\right]\phi_n(z)&=&\hbar \omega^c_{n\mathbf{0}}\phi_n(z),
\end{eqnarray}
where the potential due to band offset $V(z)$ and the effective mass $m^*(z)$ are piecewise constant.
Solution of \Eq{BDD} can then be numerically calculated enforcing the continuity of $\phi_n(z)$ and $\frac{1}{m^*(z)}\frac{\partial}{\partial z}\phi_n(z)$ at the interfaces \cite{Bastard}. 

\section{Theory of bound-to-continuum strong coupling}
The Power-Zienau-Wooley Hamiltonian, initially introduced to describe multisubband polaritons \cite{Todorov12,Todorov15}, reads
\begin{eqnarray}
\label{HPZW}
H&=&\sum_{\mathbf{q}}\left[\hbar\omega^a_q a^{\dagger}_{\mathbf{q}}a_{\mathbf{q}}+\sum_{\alpha} \omega^b_{\alpha} b^{\dagger}_{\alpha \mathbf{q}}b_{\alpha \mathbf{q}}
+ \sum_{\alpha}\frac{\hbar\omega_{P\alpha} d_{\alpha}}{2} \sqrt{\frac{\omega^a_{q}}{L_c I_{\alpha\alpha}\omega^b_{\alpha}}} (a^{\dagger}_{\mathbf{-q}}+a_{\mathbf{q}})
(b^{\dagger}_{\alpha \mathbf{q}}+b_{\alpha\mathbf{-q}})\right.\\ &&\left.+\sum_{\alpha,\beta}
\frac{I_{\alpha\beta}}{\sqrt{I_{\alpha\alpha}I_{\beta\beta}}}
\frac{\hbar \omega_{P\alpha}\omega_{P\beta}}{4\sqrt{\omega^b_{\alpha}\omega^b_{\beta}}}
(b^{\dagger}_{\alpha \mathbf{q}}+b_{\alpha\mathbf{-q}})(b_{\beta\mathbf{-q}}^{\dagger}+b_{\beta\mathbf{q}})\right].\nonumber
\end{eqnarray}
In \Eq{HPZW}, for a in-plane wavevector $\mathbf{q}$, the operator $a_{\mathbf{q}}^{\dagger}$ describes a photon with frequency $\omega^a_q$ and $b^{\dagger}_{\alpha\mathbf{q}}$, $\alpha\equiv(nm)$, a collective electronic transition.  Due to the quasi-parabolic subbands dispersion and the smallness of the photonic momentum we will always neglect the transition dispersion \cite{Ciuti05}
\begin{eqnarray}
\omega^c_{n\lvert \mathbf{k+q}\rvert}-\omega^c_{mk}\simeq
\omega^c_{n0}-\omega^c_{m0},
\end{eqnarray}
leading to a dispersionless transition frequency
\begin{eqnarray}
\omega^b_{(nm)}=\omega^c_{n0}-\omega^c_{m0}.
\end{eqnarray}

The first two terms of \Eq{HPZW} describe the free photonic and excitonic fields, the third the dipolar interaction between light and matter, the fourth the dipole-dipole interaction term, often named $P^2$ in a parallel with the $A^2$ term present in the Coulomb gauge.
The remaining parameters of the Hamiltonian are $L_c$, the effective length of the photonic cavity, the plasma frequency for the transition $\alpha$
\begin{eqnarray}
\label{wp}
\omega_{P\alpha}^2&=&\frac{I_{\alpha \alpha}\hbar e^2 N_{2DEG}}{2m^{*2}_W \epsilon_0\epsilon_r \omega^b_{\alpha}},
\end{eqnarray}
and its dipole moment 
\begin{eqnarray}
\label{d}
d_{(nm)}&=&\int  \left[ \bar{\phi}_{n}(z)\partial_z \phi_{m}(z)- \phi_{m}(z)\partial_z \bar{\phi}_{n}(z)  \right]dz.
\end{eqnarray}
In \Eq{wp} the term $I_{\alpha \alpha}$ is an overlap integral between the currents generated by different transitions 
\begin{eqnarray}
\label{I}
I_{(nm)(n'm')}&=&\int \left[ \bar{\phi}_{n}(z)\partial_z \phi_{m}(z)- \phi_{m}(z)\partial_z \bar{\phi}_{n}(z)  \right]\left[ \bar{\phi}_{n'}(z)\partial_z \phi_{m'}(z)- \phi_{m'}(z)\partial_z \bar{\phi}_{n'}(z)  \right]dz.
\end{eqnarray}
Note that the effective medium approximation corresponds to assume a factorised form for the overlap integral,
\begin{eqnarray}
I_{\alpha\beta}&=&\sqrt{I_{\alpha\alpha}I_{\beta\beta}}.
\end{eqnarray}
In order to derive \Eqs{wp}{I} we specialised the system to the case in which there is a single bound state per quantum well, the quantum wells are identical, and their respective bound levels can be considered quasi-degenerate. In this case each {\it bare} electronic bound state $m$ has the same population density $N_{2DEG}$, we can then neglect intersubband bound-to-bound transitions and consider dielectric parameters inside the quantum well.

We start considering at first the matter part of the Hamiltonian
\begin{eqnarray}
H_M&=&\sum_{\mathbf{q}}\left[\sum_{\alpha} \omega^b_{\alpha} b^{\dagger}_{\alpha \mathbf{q}}b_{\alpha \mathbf{q}}
+\sum_{\alpha,\beta}
\frac{I_{\alpha\beta}}{\sqrt{I_{\alpha\alpha}I_{\beta\beta}}}
\frac{\hbar \omega_{P\alpha}\omega_{P\beta}}{4\sqrt{\omega^b_{\alpha}\omega^b_{\beta}}}
(b^{\dagger}_{\alpha \mathbf{q}}+b_{\alpha\mathbf{-q}})(b_{\beta\mathbf{-q}}^{\dagger}+b_{\beta\mathbf{q}})\right],
\end{eqnarray}
which can be diagonalised in terms of multisubband plasmon operators as
\begin{eqnarray}
H_M&=&\sum_{\alpha\mathbf{q}} \hbar\omega_{\alpha}^pp_{\alpha\mathbf{q}}^{\dagger}p_{\alpha\mathbf{q}},
\end{eqnarray}
where the collective transition operators between single-particle states can be expressed as linear superpositions of multisubband plasmons
\begin{eqnarray}
(b^{\dagger}_{\alpha \mathbf{q}}+b_{\alpha\mathbf{-q}})&=&\sum_{\beta}h_{\alpha \beta}(p_{\beta \mathbf{q}}^{\dagger}+p_{\beta \mathbf{-q}}).
\end{eqnarray}
Introducing the Coulomb-renormalised coupling coefficient
\begin{eqnarray}
\Xi_{\alpha q}&=&\sqrt{\frac{\omega^a_{q}}{L_c }}\left[\sum_{\beta}h_{\beta \alpha }\frac{\omega_{P\beta} d_{\beta}}{\sqrt{I_{\beta\beta}\omega^b_{\beta}}}\right],
\end{eqnarray}
the full Hamiltonian in \Eq{HPZW} can be put in the form
\begin{eqnarray}
H&=&\sum_{\mathbf{q}}\left[\hbar\omega^a_q a^{\dagger}_{\mathbf{q}}a_{\mathbf{q}}+\sum_{\alpha} \hbar\omega^p_{\alpha}p_{\alpha\mathbf{q}}^{\dagger}p_{\alpha\mathbf{q}}
+ 
\sum_\alpha \frac{\hbar \Xi_{\alpha q}}{2}
 (a^{\dagger}_{\mathbf{-q}}+a_{\mathbf{q}})
(p_{\alpha\mathbf{q}}^{\dagger}+p_{\alpha\mathbf{-q}})\right],
\end{eqnarray}
and diagonalised in terms of bosonic hybrid light-matter operators
\begin{eqnarray}
\label{polopSup}
d_{s\mathbf{q}}=x_{sq} a_{\mathbf{q}}+z_{sq} a^{\dagger}_{\mathbf{-q}}+
\sum_{\alpha} \left[ y_{s \alpha q}p_{\alpha \mathbf{q}}+w_{s \alpha q}p_{\alpha\mathbf{-q}}^{\dagger}\right],
\end{eqnarray}
with frequencies $\omega^d_{sq}$, leading to the eigenvalue equation
\begin{eqnarray}
\label{eigenwsupp}
\frac{\omega^a_{q}}{\omega_{q}^{a2}-\omega^{d2}_{sq}}\sum_{\alpha}
\frac{\lvert \Xi_{\alpha q}\rvert^2\omega^p_{\alpha}}{\omega_{\alpha}^{p2}-\omega^{d2}_{sq}}&=&1.
\end{eqnarray}

\section{Calculation of the strongly coupled electronic wavefunctions}
We aim to calculate the difference in the electronic population between the ground state and  a polaritonic $d_{s\mathbf{q}}$ state
\begin{eqnarray}
\Delta N_{s}(z)&=&\bra{G}d_{s\mathbf{q}} N(z)d^{\dagger}_{s\mathbf{q}}\ket{G}-\bra{G}N(z) \ket{G}.
\end{eqnarray}
We start by expressing the hybrid light-matter operators in terms of the bare single electron transitions
\begin{eqnarray}
d_{s\mathbf{q}}&=&x_{sq} a_{\mathbf{q}}+z_{sq} a^{\dagger}_{\mathbf{-q}}+
\sum_{\alpha} \left[ \tilde{y}_{s \alpha q}b_{\alpha \mathbf{q}}+\tilde{w}_{s \alpha q}b_{\alpha \mathbf{-q}}^{\dagger}\right],
\end{eqnarray} 
using then their definition in terms of electronic states
\begin{eqnarray}
b^{\dagger}_{\alpha\mathbf{q}}&=&\frac{1}{\sqrt{N}}\sum_{\mathbf{k}}c_{n\mathbf{k+q}}^{\dagger}c_{m\mathbf{k}},
\end{eqnarray}
to calculate the commutators
\begin{eqnarray}
\left[a_{\mathbf{q}}, N(z)
 \right]&=& \left[a^{\dagger}_{-\mathbf{q}}, N(z)
 \right]=0,\\
 \left[b_{(nm)\mathbf{q}}, N(z)
 \right]&=&
 \sum_{n'>n_{QW}}  \phi_n(z) \phi_{n'}(z) b_{(n'm)\mathbf{q}}-\sum_{m'\leq n_{QW}} \phi_m(z)\phi_{m'}(z) b_{(nm')\mathbf{q}},\\
 \left[b^{\dagger}_{(nm)\mathbf{-q}}, N(z)
 \right]&=&\sum_{m'\leq n_{QW}} \phi_m(z)\phi_{m'}(z) b^{\dagger}_{(nm')\mathbf{-q}}-
 \sum_{n'>n_{QW}}  \phi_n(z) \phi_{n'}(z) b^{\dagger}_{(n'm)\mathbf{-q}},
\end{eqnarray}
where $n_{QW}$ is the number of quantum wells equal to the number of bare electronic bound states.
Exploiting those commutators we are able to calculate
\begin{eqnarray}
\Delta N_{s\mathbf{q}}(z)&=&\sum_{nn'> n_{QW}}   \phi_n(z) \phi_{n'}(z) M_{Esq}^{nn'}
-\sum_{mm'\leq n_{QW}}   \phi_m(z) \phi_{m'}(z) M_{Gsq}^{mm'},
\end{eqnarray}
with
\begin{eqnarray}
M_{Esq}^{nn'}&=&\sum_{m\leq n_{QW}} \left[ \tilde{y}_{s(nm)q}\tilde{y}_{s(n'm)q}+\tilde{w}_{s(nm)q}\tilde{w}_{s(n'm)q}\right],\\
M_{Gsq}^{mm'}&=&\sum_{n> n_{QW}} \left[ \tilde{y}_{s(nm)q}\tilde{y}_{s(nm')q}+\tilde{w}_{s(nm)q}\tilde{w}_{s(nm')q}\right].
\end{eqnarray}
Noticing that
\begin{eqnarray}
\int dz \sum_{nn'> n_{QW}}   \phi_n(z) \phi_{n'}(z) M_{Esq}^{nn'}
=\int dz \sum_{mm'\leq n_{QW}}   \phi_m(z) \phi_{m'}(z) M_{Gsq}^{mm'}
=\sum_{\scriptsize{\begin{array}{c}n>n_{QW}\\ m\leq n_{QW}\end{array}}} \left[ \tilde{y}_{s(nm)q}^2+\tilde{w}_{s(nm)q}^2\right]
\end{eqnarray}
as required from charge conservation, we can thus write 
\begin{eqnarray}
\Delta N_{s\mathbf{q}}(z)&=&P_{sq}\left[\lvert \psi^e_{sq}(z) \rvert^2-\lvert \psi^g_{sq}(z) \rvert^2\right],\end{eqnarray}
where the matter component weight is given by
\begin{eqnarray}
P_{sq}&=&\sum_{\scriptsize{\begin{array}{c}n>n_{QW}\\ m\leq n_{QW}\end{array}}} \left[ \tilde{y}_{s(nm)q}^2+\tilde{w}_{s(nm)q}^2\right],
\end{eqnarray}
and the normalised ground and excited electron densities read
\begin{eqnarray}
\lvert \psi^g_{sq}(z) \rvert^2&=&\frac{1}{P_{sq}}
 \sum_{mm'\leq n_{QW}}   \phi_m(z) \phi_{m'}(z) M_{Gsq}^{mm'},\\
\lvert \psi^e_{sq}(z) \rvert^2&=&\frac{1}{P_{sq}}\sum_{nn'> n_{QW}}   \phi_n(z) \phi_{n'}(z) M_{Esq}^{nn'}.
\end{eqnarray}
}

\end{document}